# The Origin of the Observed Raman Peaks in α-MnTe

Nurul Azam,[1,2,†] Syed Mohammad Shahed,[1,2] Liam T. Schmidt,[1,2] Sara Bey,[3] Oksana Yastrubchak,[4] Maria F Munoz,[5] Riccardo Torsi,[5] Thi Hai Yen Pham,[6,7] Dushyanthini Balasundaram,[6,7] Resham Babu Regami,[3,10] Wentao Liang,[8] Imrankhan Mulani,[1,2,9] Matthew Matzelle,[1,2] Vineet Kumar Sharma,[9] Sougata Mardanya,[9] Sugata Chowdhury,[9] Nirmal Ghimire,[3,10] Patrick M. Vora,[6,7] Angela R. Hight Walker,[5] Xinyu Liu,[3] Badih A. Assaf,[3] Arun Bansil,[1,2] Alberto De la Torre,[1,2] Swastik Kar[1,2,11,‡]

[1]*Department of Physics, Northeastern University, Boston, MA 02115, USA.*
[2]*Quantum Materials and Sensing Institute, Northeastern University Innovation Campus, Burlington, MA 01803, USA.*
[3]*Department of Physics and Astronomy, University of Notre Dame, Notre Dame IN, 46556, USA.*
[4]*V. E. Lashkaryov Institute of Semiconductor Physics, National Academy of Sciences of Ukraine, U-03028 Kyiv, Ukraine.*
[5] *Quantum Measurement Division, National Institute of Standards and Technology, Gaithersburg, MD, USA*
[6]*Department of Physics and Astronomy, George Mason University, Fairfax, VA 22030, USA*
[7]*Quantum Science and Engineering Center, George Mason University, Fairfax, VA 22030, USA*
[8]*Kostas Advanced Nano-Characterization Facility, Northeastern University Innovation Campus, Burlington, MA, USA.*
[9]*Department of Physics and Astrophysics, Howard University, Washington, DC 20059, USA.*
[10]*Stavropoulos Center for Quantum Matter, University of Notre Dame, Notre Dame IN, 46556, USA.*
[11]*Department of Chemical Engineering, Northeastern University, Boston 02115, USA.*

*Corresponding authors: †Nurul Azam (m.azam@northeastern.edu), ‡Swastik Kar (s.kar@northeastern.edu)*

**Abstract**

The Raman spectrum of the room-temperature altermagnet α-MnTe is reported to contain unassigned peaks at 120(3) $cm^{-1}$ and 140(3) $cm^{-1}$, absent from the predicted phonon spectrum of the material. This has generated considerable debate within the altermagnet community and necessitates urgent resolution. This work establishes that these peaks, together with the 90(5) $cm^{-1}$ peak that matches a theoretically predicted mode, are all extrinsic, originating from elemental tellurium formed during air exposure of the surface. Raman spectra of MBE-grown thin films show that these peaks closely match those of elemental tellurium, emerge soon after air exposure, and are absent in $AlO_x$-capped films. X-ray photoelectron spectroscopy shows that air exposure breaks Mn–Te bonds and oxidizes Mn within a minute. Cross-sectional scanning transmission electron microscopy with energy-dispersive X-ray spectroscopy reveals that the oxidation leads to Mn out-diffusion, forming a few-nanometer-thick oxide layer above a buried, Te-enriched region, likely responsible for the anomalous Raman peaks. Additionally, no sample exhibited the 175 $cm^{-1}$ Raman peak which is commonly attributed to $MnTe_2$. The aggressive surface oxidation and associated elemental Te formation in MnTe have important implications for any surface-sensitive and optical characterization of MnTe (and other similar Te-containing materials) involving even the briefest air exposure.

## Introduction

In recent times, semiconducting α-MnTe (space group *P6₃/mmc*, Néel temperature $T_N \approx 307$ K) has received enormous attention as a prototypical, room-temperature candidate that hosts altermagnetism, a third elementary class of magnetism, distinct from both ferromagnetism and antiferromagnetism [1],[2],[3]. Altermagnets host collinear spin arrangements with zero net magnetization in which spin-opposite sublattices are related by a rotation, instead of being connected by a translation or inversion as in conventional antiferromagnets [1], [2], [3]. This symmetry distinction leads to a momentum-dependent spin splitting of the electronic and magnetic degrees of freedom in the absence of net magnetization [4], making altermagnets unique candidates for spintronic technologies [5]. In MnTe, the Mn moments lie ferromagnetically aligned in the ab-plane, but are antiferromagnetically aligned between the planes. A nonsymmorphic sixfold screw rotation relates the face-sharing $MnTe_6$ octahedra in different spin sublattices [2],[6], thus fulfilling the symmetry conditions for altermagnetism. These symmetry constraints manifest in a large anomalous Hall effect, exchange-induced band splitting, and chiral magnon bands that are unique to altermagnets [7], [8], [9], [10], [11],[12],[13], [14]. Despite the rapid progress in our understanding of the microscopic origin of magnetic, transport, and other properties, unambiguously establishing the altermagnetic classification of MnTe requires experimental techniques directly sensitive to both the structural and magnetic point groups.

Raman spectroscopy has been widely used in magnetic materials to characterize local symmetries, phononic and spin degrees of freedom, and their coupling [15] [16], [17],[18],[19],[20]. In altermagnets, the phonon spectrum offers a direct route to adjudicating whether a candidate compound retains its parent symmetry or hosts a subtle structural distortion of the kind that would modify the spin-group description. [21], [22] In MnTe the Raman spectrum has proven remarkably difficult to interpret. The reported spectrum is dominated by two intense peaks at 120(3) $cm^{-1}$ and 140(3) $cm^{-1}$, which have been widely accepted as intrinsic phonon modes of α-MnTe [21], [23], [24], [25] despite not corresponding to Γ-point optical phonons of the ideal structure [22]. Two weaker peaks near 95 $cm^{-1}$ have been attributed to a single MnTe $E_{2g}$ Raman-active mode predicted by DFT (see SI section 1.1 and references [24] and [22]), yet local strain or the global loss of six-fold symmetry had to be invoked to account for the experimentally observed splitting. A separate feature reported near 175 $cm^{-1}$ has been interpreted as "leakage" of an otherwise silent mode activated by an inversion-symmetry-breaking distortion [21], proposed to be an electronic (plasmonic) excitation enabled by hole self-doping [22], and, alternatively, shown to originate entirely from $A_g$ and $T_g$ modes from a secondary $MnTe_2$ phase absent in phase-pure crystals [23], [26], [27]. Discrepancies in the Raman spectrum of tellurides are not unique to MnTe, but have complicated the interpretation of a wide array of Te-based chalcogenides [16], [28], [29], owing in part to elemental Te modes that lie in the same energy range [30]. Beyond assigning the observed modes, resolving the origin of the Raman response in MnTe carries direct implications for whether the material belongs to an altermagnetic symmetry class.

Here, we show that the three anomalous Raman peaks observed near 90(5) $cm^{-1}$, 120(3) $cm^{-1}$ and 140(3) $cm^{-1}$ in α-MnTe are extrinsic and arise due to the segregation of elemental tellurium generated during surface oxidation. By tracking the time evolution of the Raman response of MnTe/ GaAs molecular beam epitaxy (MBE)-grown thin films, we find that these peaks emerge within approximately 26 minutes of ambient exposure and subsequently intensify over time, a process that gets markedly accelerated by continuous laser exposure. Moreover, these peaks are absent in the Raman spectra of air-exposed $AlO_x$ capped thin films. Combining exposure-dependent X-ray photoelectron spectroscopy (XPS) with cross-sectional scanning transmission electron microscopy with energy-dispersive X-ray spectroscopy (STEM-EDS), we establish that ambient oxygen aggressively breaks the Mn–Te bonds at the surface and rapidly oxidizes Mn. As a result, Mn diffuses outward to form a few-nanometer-thick oxide and hydroxide overlayer, leaving behind a Mn-depleted and Te-rich under-layer. We thus attribute the anomalous Raman modes to elemental tellurium residing beneath the surface oxide. These findings demonstrate that even the most nominal degree of air-exposure can lead to dramatic surface degradation in MnTe, with direct implications for surface-sensitive techniques and other telluride samples.

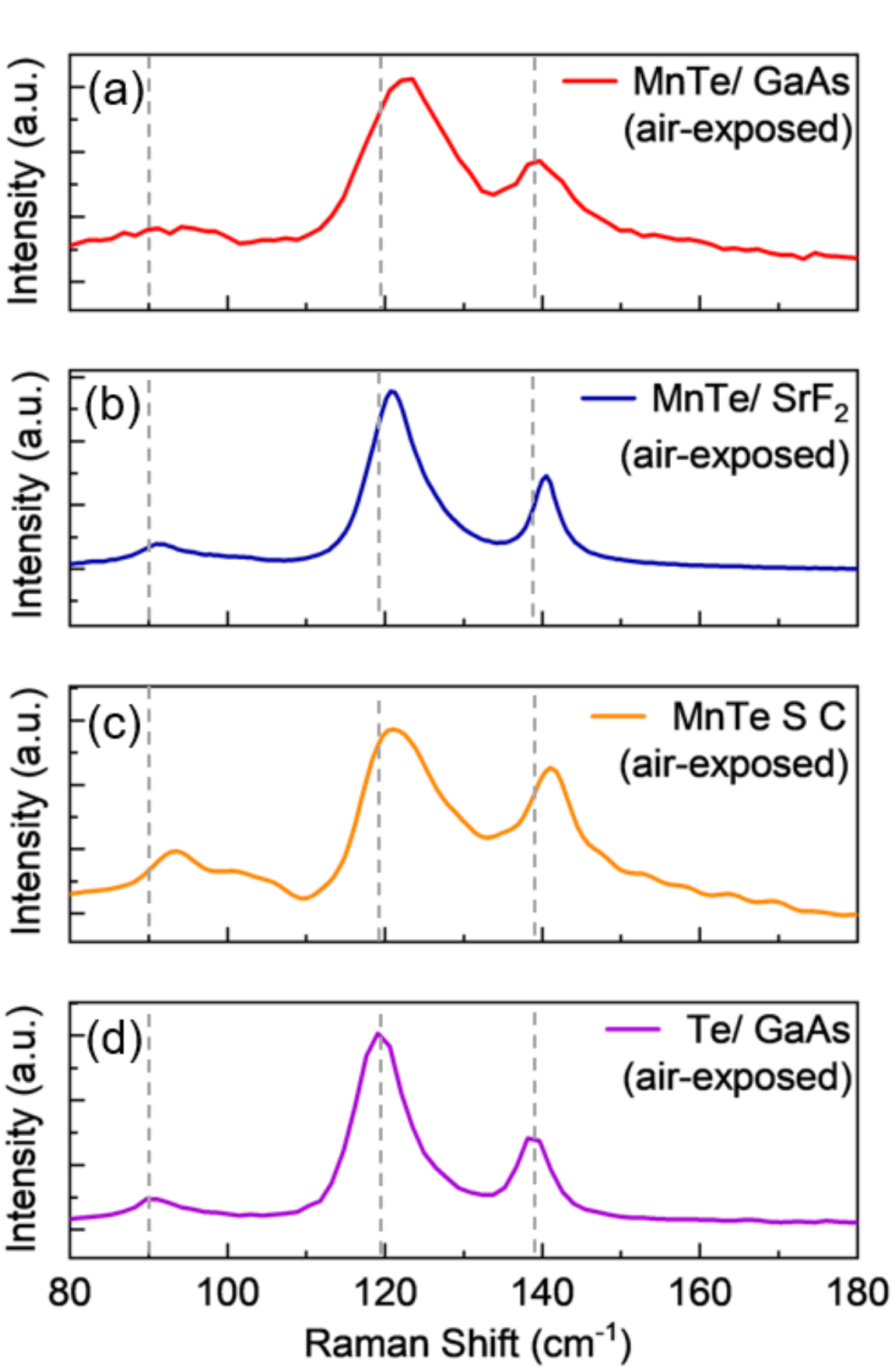

**Figure 1. Comparison of Raman spectra from air-exposed MnTe and elemental Te collected with 532 nm excitation.** Normalized Raman spectra between 80 $cm^{-1}$ and 180 $cm^{-1}$ are shown for (a) MnTe film on GaAs (111), (b) MnTe film on $SrF_2$ (111), (c) bulk single crystal (S C), and (d) Te/GaAs (111) thin-film reference. Dashed vertical lines mark the features near 90(5) $cm^{-1}$, 120(3) $cm^{-1}$, and 140(3) $cm^{-1}$.

## Results and Discussion

We first demonstrate the ubiquitous occurrence of the Raman features across different forms of MnTe (see Methods for details of the respective growth conditions). Figure 1 compares the Raman response of air-exposed MnTe thin-film and bulk alongside a pure Te thin-film reference, all measured using 532 nm laser excitation. MnTe films grown by MBE on GaAs (111) (Figure 1a) and $SrF_2$ (111) (Figure 1b), together with bulk single-crystal MnTe (Figure 1c), show reproducible features near 120(3) $cm^{-1}$ and 140(3) $cm^{-1}$ and a weaker band near 90(5) $cm^{-1}$. These features agree with previously reported but unassigned Raman modes of MnTe.[23] We note that the mode at 90(5) cm⁻¹ is broad and could be accounted for by two independent peaks. Data are individually normalized to the peak intensity of their respective ≈ 120(3) $cm^{-1}$ mode. Although the relative intensities and linewidths vary among the MnTe samples, the observed peak positions consistently agree closely with the characteristic $E_1$, $A_1$, and $E_2$ modes of trigonal Te (Figure 1d), as well as with anomalous Raman modes observed in other tellurides [28], [29], [31]. Angle-resolved Raman of an air-exposed MnTe/ GaAs sample shows agreement with the assigned symmetries of the trigonal Te (Figure S2). However, while surface oxidation of α-MnTe/ GaAs has been shown to produce an amorphous $MnO_x$ surface layer together with a Te-rich, Mn-deficient subsurface region [32], the formation of elemental Te clusters, their connection to the Raman features, and resulting dynamic contribution to the evolving Raman response have not yet been established. To understand this, we first discuss the *in-situ* characterization of chemically pristine MBE-grown MnTe/GaAs films.

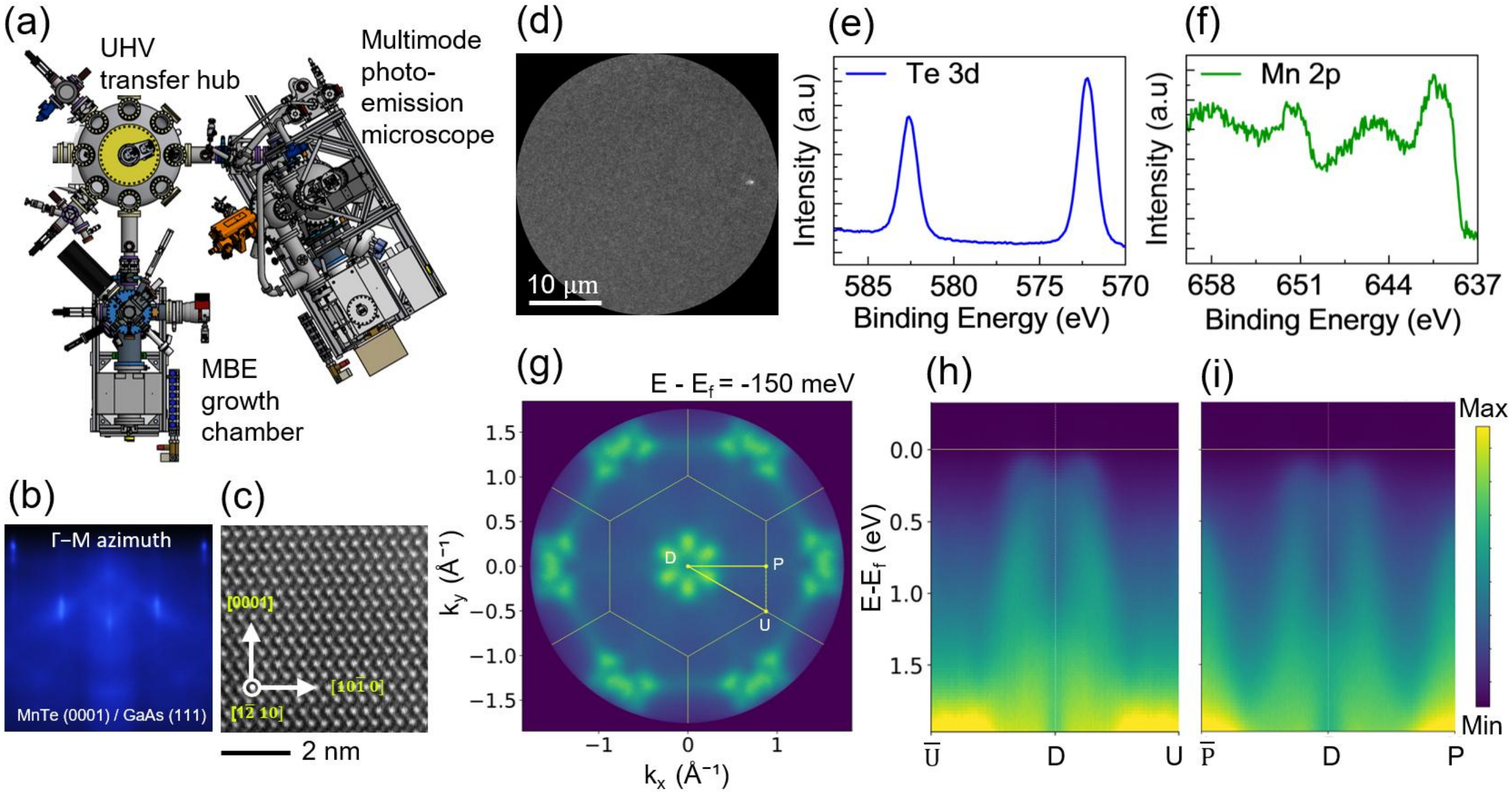

**Figure 2. Vacuum-integrated growth and in-situ surface characterization, with complementary ex-situ structural analysis, of epitaxial MnTe (0001) on GaAs (111).** (a) Schematic of the ultrahigh-vacuum platform connecting the molecular beam epitaxy (MBE) chamber to the multimode photoemission microscope through a vacuum-transfer manifold, enabling characterization of the as-grown MnTe surface without air exposure. (b) RHEED patterns of the MnTe (0001) film along the Γ-M azimuth. (c) complementary ex-situ atomic-resolution STEM, XRD, and SAED measurements were performed after removal from vacuum (see SI, Figure S3 for XRD and SAED). Although the film was exposed to air, the STEM image probes an interior region unaffected by surface oxidation. Together, these measurements confirm the high crystalline quality of MBE grown MnTe film. (d) photoemission electron microscopy (PEEM) image of the as-grown surface over a 32 μm field of view. (e,f) core-level Te 3d and Mn 2p XPS spectra for a pristine film. (g) Momentum microscopy showing a constant-binding-energy momentum map acquired at $E - E_f = -150\ m$eV below the Fermi level, measured at ≈20 K temperature. (h,i) Momentum-resolved photoemission dispersions measured along $\overline{\text{U}}$-D-U and $\overline{\text{P}}$-D-P, respectively, revealing the valence-band structure of the MnTe film.

Figure 2a shows a schematic of the UHV setup combining our MBE system with a multi-mode photoemission microscope. Figure 2b shows a characteristic reflection high-energy electron diffraction (RHEED) pattern along the Γ-M azimuth, confirming the epitaxial growth of the α-MnTe (0001)/ GaAs (111) films. Complementary atomic-resolution STEM imaging (Figure 2c; also see SI, Figure S3), performed *ex-situ* but from deep inside the film interior, further confirms that the synthesized chemically pristine MnTe possesses high-quality long-range crystalline order. The pristine samples are transferred into our nanoscale electron spectroscopy for chemical analysis (Nano-ESCA) system across a UHV-connected manifold to perform multi-mode photoemission characterizations while avoiding air exposure. Photoemission electron microscopy (PEEM) of the pristine sample (Figure 2d; 32 µm field of view) was acquired at a kinetic energy of 4.8 eV, just above the photoemission threshold, using a mercury lamp. The uniform PEEM intensity from our α-MnTe films, with small grain contrast, indicates a constant work function and a smooth surface, confirming the structural and chemical uniformity of the epitaxial layer. Core-level XPS spectra of Te 3d and Mn 2p (Figure 2e and Figure 2f) establish the chemical-state line shapes of MnTe prior to oxidation. These spectra therefore serve as the zero-exposure reference for tracking subsequent changes in the Mn and Te core levels during controlled air exposure, as discussed in detail in Figure 3**.** The electronic structure of the pristine film at ≈20 K temperature was further examined by momentum microscopy using a He I discharge lamp ($\hbar\omega$ = 21.22 eV). The constant-binding-energy map at $E - E_f = -150\ m$eV below the Fermi level (Figure 2g**)** exhibits a sixfold momentum-space pattern consistent with the hexagonal symmetry of the α-MnTe (0001) surface Brillouin zone. In Figure 2h and Figure 2i, we show energy *vs.* momentum cuts along the high-symmetry directions $\overline{\text{U}}$-D-U and $\overline{\text{P}}$-D-P, respectively. Our data establish the characteristic near-$E_F$ Te 5p states, consistent with previous reports in the literature [7], [30], [31],[32],[33],[34]. Taken together, our RHEED, STEM, PEEM, XPS, and momentum microscopy data establishes the high quality of our chemically pristine samples. We now turn our attention to the effects of air exposure on the surface of α-MnTe. Figure 3 shows the changes in the O1s, Mn $2p_{3/2}$, and Te $3d_{5/2}$ core level XPS spectra as a pristine α-MnTe/GaAs sample is systematically exposed to atmospheric conditions. The 0-minute exposure corresponds to the as-grown sample measured in UHV.

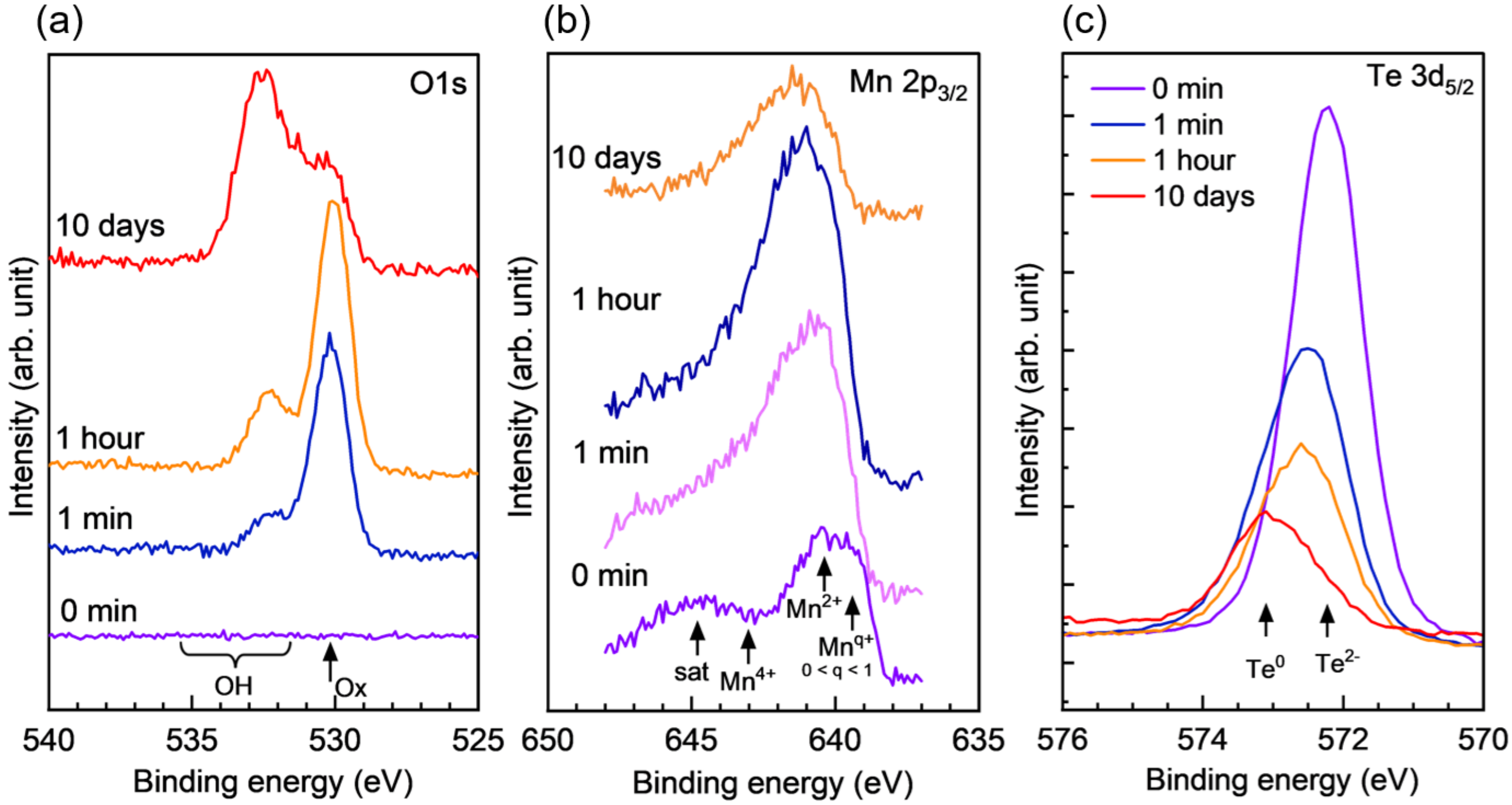

**Figure 3. Evolution of the O1s, Mn $2p_{3/2}$, and Te $3d_{5/2}$ core-level XPS spectra with increasing air exposure, 0 min (as grown) to 10 days, revealing the surface oxidation dynamics of MnTe.** (a) O1s peaks at ≈530.0 eV and ≈532.0 eV emerge upon air exposure, indicating Mn-oxide/hydroxide formation. Spectra are vertically offset for clarity. (b) Mn $2p_{3/2}$ peaks ($Mn^{q+}$, $Mn^{2+}$) shift to higher binding energy and the ≈ 645.0 eV satellite (Te to Mn charge transfer) rapidly disappears. Spectra are vertically offset for clarity. (c) The $Te^{2-}$ peak at ≈ 572.2 eV shifts to ≈ 573.0 eV ($Te^0$) with apparent decreasing intensity, indicating oxidation of MnTe and segregation of elemental Te. All peak-position assignments are based on the apparent peak positions and are supported by previous reports, as discussed in the main text.

The O1s spectrum (Figure 3a) in the pristine sample confirms that the concentration of surface oxygen is below the XPS detection limit. The appearance of a strong oxygen peak within a 1-minute exposure to ambient air reflects the dramatic rate of surface oxidation. The O1s region evolves substantially with increased exposure, with peaks initially emerging at ≈530.0 eV and over time, getting dominated by a ≈ 532.0 eV to ≈ 534.0 eV broad peak that closely matches reported XPS values for metal oxides, hydroxides, and surface adsorbed water respectively,[35], [36], [37], [38] confirming significant and rapid degradation of the pristine α-MnTe surface. The Mn $2p_{3/2}$ spectrum in Figure 3b for the unexposed film shows peaks at ≈ 639.7 eV, ≈ 640.5 eV, and ≈ 645.0 eV, which correspond to $Mn^{q+}$ ($0 < q < 1$), $Mn^{2+}$, and a satellite peak respectively, consistent with the reported observations for MnTe [39], [40], [41]. The satellite peak at ≈ 645.0 eV originates from ligand-to-metal charge transfer (Te to unfilled Mn d-orbital) [39], [40], [42], [43]. We observed mainly two changes in the Mn $2p_{3/2}$ region after air exposure: (1) gradual shift of the ≈ 639.7 eV and ≈ 640.5 eV peaks to higher binding energy, and (2) disappearance of the ≈ 645.0 eV satellite peak. The binding energy shift is consistent with previous reports attributing higher-binding-energy peaks to Mn oxides in various oxidation states [35], [40]. The rapid disappearance of the satellite peak, within just a minute of air exposure, indicates that Mn-O bond formation weakens or breaks the Mn-Te bonds, which should also alter the Te oxidation states. Figure 3c confirms the corresponding changes in Te oxidation state, consistent with those observed for the Mn $2p_{3/2}$ peak. The Te $3d_{5/2}$ spectrum of the pristine sample shows a peak at ≈ 572.2 eV, attributed to the $Te^{2-}$ state in MnTe [39], [44]. With increasing air exposure (one minute to 10 days), this peak gradually shifts to ≈ 573.0 eV, accompanied by an apparent decrease in intensity or peak area (quantitative analysis of peak area changes is beyond the scope of this study). We attribute the ≈ 573.0 eV component to elemental Te ($Te^0$), confirmed by a separate Te 3d XPS measurement on an MBE-grown Te film on GaAs (Supplementary Figure S4). Intermediate components between the $Te^{2-}$ and $Te^0$ states may correspond to Te atoms loosely bonded to Mn during the transition. To summarize, we observed several key trends here with increasing air exposure: (1) rapid Mn oxidation, evidenced by rapid disappearance of the ≈ 645.0 eV satellite peak and emergence of the O 1s peak at 1-minute exposure, (2) no apparent drop in Mn $2p_{3/2}$ peak area, and (3) conversion of $Te^{2-}$ to $Te^0$, accompanied by a substantial drop in Te $3d_{5/2}$ peak area. These observations suggest that the MnTe surface is heavily oxidized within the 1-minute of air exposure, as oxygen weakens or breaks Mn-Te bonds and forms Mn-$O_x$ bonds. The decreasing Te peak intensity despite retention of the strong Mn peak suggests the possibility that upon exposure to air, the bond-broken Mn atoms diffuse toward and possibly above the surface, promoting growth of a Mn-oxide layer [32] and segregation of elemental Te buried beneath this layer. The decrease in Te $3d_{5/2}$ peak area is due to attenuation of the Te signal as the overlying Mn-oxide layer thickens. To confirm this hypothesis, we performed detailed element-resolved, near-to-top surface cross-sectional STEM-EDS analysis of an α-MnTe/GaAs film that was exposed to air for several weeks (Figure 4).

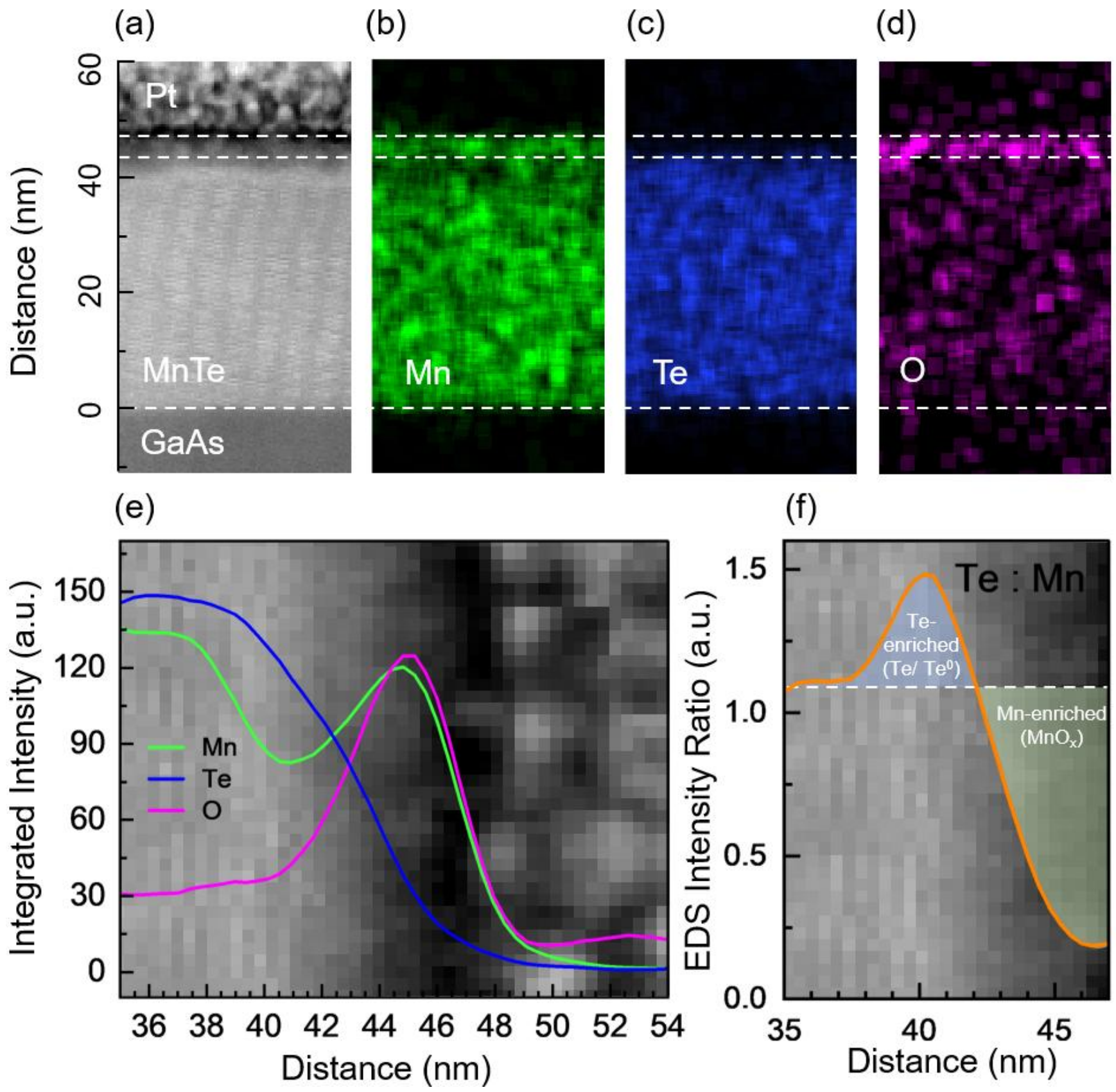

**Figure 4. Cross-sectional STEM-EDS analysis of an MnTe/GaAs film after few weeks of air exposure.** (a) Cross-sectional STEM image and (b-d) show corresponding EDS maps of Mn, Te, and O EDS maps across the GaAs/MnTe/$MnO_x$/Pt stack. Horizontal dashed lines mark the GaAs/MnTe interface and the boundaries of the approximately 3 nm to 5 nm thick $MnO_x$ layer formed after few weeks of air exposure. Mn and Te are distributed throughout the underlying MnTe film, whereas O is concentrated within the near-surface region. (e) Integrated Mn, Te, and O EDS profiles across the near-surface region, overlaid on the corresponding STEM image. (f) Relative Te:Mn EDS intensity ratio across the MnTe-to-$MnO_x$ transition. The dashed line represents the average ratio within the underlying MnTe region.

The STEM image (Figure 4a) shows an approximately 47 nm thick film between the GaAs substrate (bottom horizontal dashed line) and the protective Pt cap (top dashed line). The protective Pt cap was deposited after air exposure and preserved the pre-existing surface during focused ion beam lamella preparation for TEM/EDS characterization. The middle-dashed line denotes the approximate position of the original MnTe film edge prior to oxidation, estimated using the Te distribution as a reference. The corresponding EDS elemental maps for Mn, Te, and O are shown in Figure 4b-d. We find that while Mn out-diffuses beyond the height denoted by the middle-dashed line towards the Pt-interface (Figure 4b), Te remains confined to the original film area (Figure 4c). Also, while O is found throughout the slab (since the cross-sectional sample was also air-exposed before insertion in the TEM tool), its highest concentration lies within the 3 nm to 5 nm region near the surface (Figure 4d). This region, between the middle and top dashed lines, coincides with the Mn out-diffused region. Our STEM-EDS data provided evidence that the few nanometer $MnO_x$ layer, chemically identified through our XPS measurements, is associated with an out-diffusion of Mn atoms from the top surface. [32] Since XPS effectively probes on the top ≈ 5 nm of the sample, this top oxide layer also explains why the observed Te XPS peaks decreased in intensity with increased air-exposure. We note that because the Pt cap was deposited only after air exposure, this oxygen-rich region represents the pre-existing oxidized MnTe surface rather than an artifact of focused ion beam preparation, as described in the Methods section. To further support this conclusion, Figure 4e shows the variation of integrated intensities of the EDS counts as a function of the sample depth, near the surface. Within the film interior, the relative integrated EDS intensities of Mn and Te remain unchanged and roughly equal, while the O intensity (from exposure during TEM sample preparations) remains low. At the edge, the Te signal decreases steadily between ≈ 40 nm and ≈ 45 nm distance. In contrast, Mn first decreases and then increases again as the O signal grows, resulting in both Mn and O intensities reaching a maximum near the 45-nm mark. Their spatial overlap, together with the pronounced Te depletion in the same region, identifies the surface layer as Mn-rich and Te-deficient, consistent with the formation of $MnO_x$. Immediately below this layer, the relative ratio of the integrated EDS counts (Te:Mn) rises above that of the underlying MnTe (Figure 4f), reaching a maximum of ~1.5 near 40 nm, indicating local Te enrichment immediately beneath the oxidation front. The Te:Mn ratio then decreases sharply to approximately 0.2 near 45 nm as the composition becomes Mn-rich within the oxide layer. Taken together, our analysis suggests that a few-nm-thick region of mostly elemental Te within a broken MnTe matrix forms concomitantly to the surface $MnO_x$ layer. We next examine the effects of air and laser exposure on MnTe films by collecting Raman spectra at different air exposure times. Compared with surface-sensitive XPS (≈5 nm), Raman spectroscopy probes substantially deeper into the film.

Figure 5a presents the evolution of Te-like peaks in the Raman spectra of MnTe as a function of time after removal from the growth chamber, to highlight its correlation with the degree of surface oxidation. The first spectrum corresponds to ≈26 min of air exposure, the time needed to vent our UHV and mount the sample into the Raman microscope. Subsequent spectra are obtained from fresh spots on the continuously air-exposed sample. Within 26 minutes of air-exposure, a weak and broad feature appears near 150(5) $cm^{-1}$, suggesting that measurable surface modification occurred even before the first *ex-situ* data acquisition. This is consistent with the timescale of $MnO_x$ formation from our XPS measurements (Figure 3). Upon further exposure, the spectra evolve with clearly discernible Te-modes near 90(5) $cm^{-1}$, 120(3) $cm^{-1}$, and 140(3) $cm^{-1}$, discussed in Figure 1, due to the precipitation of Te below the $MnO_x$ layer. This evolution is consistent both with the oxidation mechanism established independently by XPS (Figure 3) and the cross-sectional STEM-EDS results (Figure 4). In addition to the Te-related modes, air-exposed MnTe exhibits weak but reproducible Raman features near 506 $cm^{-1}$ and 537 $cm^{-1}$ that are absent in both capped MnTe and the elemental Te reference (see SI, Figure S5). These features fall within the Mn-O vibrational range reported for manganese oxides [45], [46], [47]. Together with the Mn-oxide surface layer independently identified by XPS and STEM-EDS, their appearance provides complementary spectroscopic evidence for surface $MnO_x$ formation during air exposure.

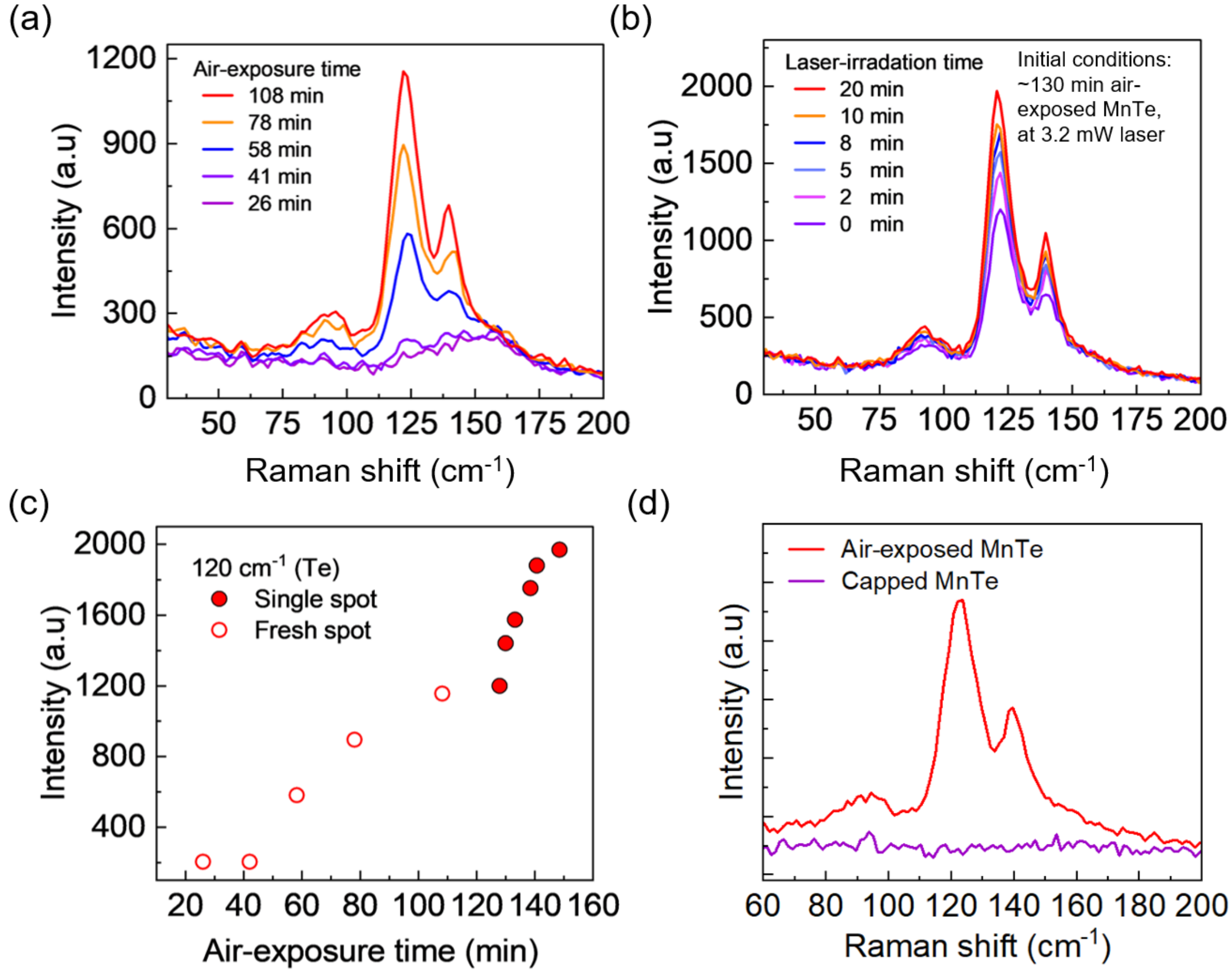

**Figure 5.** Raman spectra collected using 532 nm excitation. (a) Time-dependent evolution of Te-related Raman features during air exposure. (b) Raman spectra acquired at a single location under continuous laser irradiation of the MnTe/GaAs film after ≈130 min of air exposure. (c) Evolution of the 120(3) $cm^{-1}$ Raman peak extracted from (a) and (b). Open symbols represent measurements acquired with minimal laser exposure at different sample locations, whereas filled symbols represent prolonged irradiation at the same location. (d) Comparison of Raman spectra from capped and uncapped MnTe films. The $AlO_x$ capping suppresses oxidation and the formation of the Te-rich region.

Figure 5b further shows that prolonged laser irradiation of a fixed location accelerates the development of these Te-related bands, suggesting that continuous laser-exposure amplifies the effect of air-exposure-induced oxidation and Te segregation. This is quantitatively depicted in Figure 5c, which shows the variation of peak intensity of the 120(3) $cm^{-1}$ peak, as a function of air exposure (open symbols) and laser exposure (filled symbols). It should be noted that the MnTe sample was exposed into air for ≈130 min already when the laser exposure experiment was performed. The progressive increase of the 120(3) $cm^{-1}$ peak intensity due to ambient exposure is accelerated by continuous laser illumination of the same spot. Identical Raman features were found in the anti-Stokes modes and laser-wavelength dependent Raman spectra (See SI, Figure S6 and Figure S7). Finally, Figure 5d compares the Raman spectra of $AlO_x$-capped and uncapped MnTe films after comparable periods of air exposure. The $AlO_x$ capping layer protects the MnTe surface from air-induced oxidation [48]. The absence of Te-like modes in the capped films is consistent with the expected Raman spectrum for pristine MnTe [23],[21], [24], [25]. Furthermore, we note that in our high-quality single crystals and MnTe films, there was no discernible peak near 175 $cm^{-1}$ confirming that this peak is not associated with either pure MnTe or elemental tellurium, and does originate from $MnTe_2$ phase impurities [23], [26].

**Conclusions**

In conclusion, the combined Raman (Figure 1 and Figure 5), time-dependent XPS (Figure 3), and cross-sectional STEM-EDS (Figure 4) results show that the anomalous Raman bands in MnTe are closely linked to rapid air-induced oxidation of Mn, and an associated release/segregation of elemental Te (from $Te^{2-}$ toward $Te^{0}$) under the Mn-O overlayer. The rapid formation of Mn-O bonds and conversion of $Te^{2-}$ toward $Te^{0}$ observed in Figure 3 are consistent with the development of a Mn-rich oxide at the outer surface and a Te-enriched region immediately beneath it. Although STEM-EDS alone cannot determine the chemical state of the accumulated Te or identify the precise Mn-oxide phase, combining it with the XPS results and the correspondence between the Raman spectra of air-exposed MnTe and elemental Te in Figure 1 and Figure 5 supports a mechanism in which preferential Mn oxidation disrupts Mn-Te bonding, excludes Te from the growing oxide, and promotes Te segregation beneath the $MnO_x$ layer. The continued growth of these Te-related peaks during air exposure, and their rapid enhancement under laser irradiation, further demonstrate that the measured Raman response is strongly influenced by Raman-active Te species rather than by the intrinsic MnTe lattice.

Although past works have argued the anomalous Raman modes in tellurides [28] could arise from trigonal Te (t-Te) nanostructures, the microscopic mechanism for how this could happen in MnTe was completely unclear, especially since a related past work [32] (and our present measurements) did not reveal any evidence of such nanostructures in high-resolution STEM images of air-exposed (but laser un-exposed) samples. Interestingly, it turns out that even a few pulses of femtosecond laser exposure can result in the formation of embedded t-Te nanocrystals in tellurite glasses [49],[50]. Taken together, these suggest that while surface oxidation of Mn results in the unbonded elemental Te to form a segregated sub-surface layer in our MnTe samples, the very act of Raman measurement drives them to nucleate and crystallize into t-Te phases – as indicated by the stark acceleration of peak-height enhancement under continuous illumination (Figure 5c). In other words, in MnTe, the oxidation of out-diffused Mn, debonding and segregation of elemental Te, and exposure to laser all contribute in tandem towards the nucleation and growth of sub-surface t-Te nanocrystals – which in turn result in the appearance and growth of the anomalous Raman peaks. We emphasize that no single technique used here would have been sufficient for deriving these conclusions, which was made possible only by the convergence of Raman, XPS, STEM-EDS, performed on samples of various types, with and without $AlO_x$-capped control, along with control Te films. This unambiguous experimental evidence of this microscopic mechanism hence not only resolves the chemical origin of the disputed peaks, but also the microscopic mechanism through which they form during Raman spectroscopy.

The rapid pace of oxidation and surface degradation has significant implications on all surface-sensitive measurements including photoemission and scanning tunneling microscopy in MnTe and other similar systems. During our investigations (data not shown), we found that samples kept under UHV ($\sim 10^{-10}$ mbar) for extended periods did not display any measurable oxidation peaks in XPS. Air-exposed samples on which Raman spectra were measured through a vacuum optical port (again, data not shown) displayed reduced or halted progress in the growth of the "Te"-like peaks. These reflect the need for research on reliable handling, capping, clean de-capping, and laser-exposure under controlled conditions as a critical need for reliable Raman analysis of MnTe and related altermagnetic chalcogenides. Our finding also plausibly explains the two anomalous broad Raman bands frequently observed in bulk and low-dimensional Te-based chalcogenides, where oxidation-induced Te segregation may generate vibrational signatures resembling those of elemental Te [28], [29], [31]. More broadly, the same aggressive Mn-Te surface chemistry is likely relevant to other functional manganese- and tellurium-based quantum materials under active study, including the topological insulator $Bi_2Te_3$ and the intrinsic magnetic topological insulator $MnBi_2Te_4$ [44], a platform for the axion-insulator state, both of which likewise combine air-sensitive, tellurium-terminated surfaces with a strong reliance on surface-sensitive optical and spectroscopic characterization. Hence, we believe that our work has highly important implications for the broader chalcogenide research community.

## Experimental Methods

### α-MnTe (0001) /GaAs (111) film growth at Northeastern

MnTe thin films were grown by molecular beam epitaxy in a Scienta Omicron Lab 10 ultrahigh-vacuum setup installed as part of the Materials Innovation Platform at the Experiential Quantum Advancement Laboratories, Northeastern University, Burlington Campus. The system was operated at a base pressure in the low-$10^{-10}$ mbar range. GaAs (111) substrates were degassed in-situ at approximately 200 °C for several hours. The substrate temperature was then increased at a controlled ramp rate of 5 °C $min^{-1}$ to 600 °C and maintained at this temperature for 1 h to thermally desorb the native oxide layer. The GaAs (111) surface exhibits a hexagonal surface lattice with $a = 3.998$ Å, compared with $a = 4.148$ Å for bulk $\alpha$-MnTe (Table S1), corresponding to a lattice mismatch of approximately 3.75%. During deposition, the substrate manipulator thermocouple was set to 420 °C, while an optical pyrometer indicated a surface temperature of approximately 480 °C. High-purity elemental Mn (5 N) and Te (6 N) were supplied from independent effusion sources manufactured by MBE-Komponenten, with the Mn and Te reservoirs operated at approximately 830 °C and 230 °C, respectively. The Te cracker zone was maintained at 1000 °C to thermally dissociate the molecular Te flux into more reactive lower-order Te species, thereby enhancing Te incorporation during MnTe growth. The Te shutter was opened before the Mn shutter and remained open throughout the deposition. The effective MnTe growth time, defined by the Mn shutter-open interval, was 60 min, and the chamber pressure during growth was approximately $3 \times 10^{-8}$ mbar. The characteristic MnTe (0001) RHEED pattern is shown along the Γ-M azimuth (Figure 2b), and a TEM cross section was acquired along the same direction (Figure 2c). X-ray diffraction (XRD) and selected-area electron diffraction (SAED) confirm the epitaxial growth of α-MnTe, yielding lattice parameters $a \approx 4.18$ Å and $c \approx 6.70$ Å, corresponding to $c/a \approx 1.60$ (see SI, Figure S3). XRD characterization was performed using a Bruker SmartLab instrument at the MIT.nano facility.

### α-MnTe film growth at Notre Dame

Epitaxial α-MnTe films were grown by molecular beam epitaxy on $SrF_2$ (111) substrates [Figure 1b]. Prior to growth, the $SrF_2$ (111) substrates were annealed at 500 °C. The films were synthesized under Te-rich conditions at a substrate temperature of 340 °C and a deposition rate of approximately 0.4 Å/s [51]. α-MnTe films were also grown on GaAs (111) substrates. Prior to growth, the GaAs (111) substrates were annealed at 570 °C to desorb the native oxide layer and were subsequently exposed to a Te flux at a substrate temperature of 600 °C. Growth was then carried out at a substrate temperature of 400 °C and a deposition rate of approximately 0.24 Å/s [32]. The capped film shown in Figure 5d was prepared by depositing elemental Al in-situ at room temperature within the same UHV cluster. Following air exposure, the Al layer oxidized to form an ~ 6-nm-thick $AlO_x$ layer covering the entire film surface.

### α-MnTe single-crystal growth at George Mason

Single crystals of MnTe were grown by the tin-flux method. Mn pieces (Thermo Scientific; 99.9%), Te shots (Thermo Scientific; 99.999%), and Sn shots (Thermo Scientific; 99.9999%) were loaded in a 5-mL aluminum oxide crucible in a molar ratio of 1:1:20. The crucible was sealed in a fused silica ampoule under vacuum and heated to 960 °C over 10 h, homogenized at 960 °C for 12 h, and then cooled to 840 °C over 100 h. After reaching 840 °C, the excess flux was decanted from the crystals using a centrifuge, leaving behind several well-faceted, shiny hexagonal single crystal of a few millimeters in length and width. Magnetic susceptibility measurements confirmed a Néel temperature of ≈ 307 K for the synthesized MnTe single crystals (Figure S8).

**Nano-ESCA measurements at Northeastern**

XPS, PEEM, and momentum microscopy characterization of as-grown MnTe (0001) /GaAs (111) film was carried out in the connected UHV chamber at ≈20 K temperature equipped with a commercially available photoemission cluster, Nano-ESCA, jointly developed by FOCUS GmbH and Scienta Omicron GmbH. A monochromated aluminum Kα X-ray source, a mercury arc UV source, and a windowless helium gas discharge lamp (Focus HIS 14 HD) VUV source were used for XPS, PEEM, and momentum microscopy measurements, respectively. The imaging unit (microchannel plate and screen) detects incoming photoelectrons and generates the images, which were recorded using a Hamamatsu Orca-Flash4.0 digital CMOS camera, C13440-20CU.

**Momentum microscopy data processing**

Constant-energy slices were acquired in 0.025 eV steps over a momentum field of view of 4.1 $Å^{-1}$ using a $1024 \times 1024$-pixel detector, with each slice accumulated as the pixel-wise sum of 200 sweeps. Periodic artifacts from the detector's grid lines appeared as discrete peaks in the two-dimensional Fourier spectrum. Each image was Fourier transformed, and a Gaussian-tapered notch mask was applied at the corresponding spatial frequencies to suppress these artifacts while minimizing modification of the physical momentum-space signal. Because the artifact frequencies remained stationary with energy, the same mask was applied to the entire data cube before inverse Fourier transformation. For visualization only, the constant-energy map at $E - E_{\mathrm{F}} = -150$ meV was symmetrized under sixfold rotational ($C_6$) symmetry about the Γ point. The Γ-point position was determined from the center of mass of the momentum-space image, after which $\mathrm{I}\left(\mathrm{k_x}, \mathrm{k_y}\right)$ was rotated in successive 60° increments and the six symmetry-equivalent images were averaged. This symmetrization was not applied to the energy-momentum ($E$-$k$) cuts. The Fermi level, $E_{\mathrm{F}}$, was calibrated by fitting the momentum-integrated Fermi edge function.

**Raman measurements at Northeastern**

Raman measurements were performed using a Renishaw inVia™ Raman spectrometer equipped with a 2400 lines $mm^{-1}$grating and a Modu-Laser Stellar-REN multiline laser system with a maximum output power of 150 mW. A 532 nm laser was used for excitation, with an incident power of 3.2 mW at the sample, as measured using a Thorlabs PM100A compact power meter. An eclipse filter with a low-frequency cutoff of approximately 30 $cm^{-1}$ enabled the acquisition of both Stokes and anti-Stokes Raman spectra. The data presented here were collected in a confocal micro-Raman configuration using a 50× microscope objective with a 0.5 N.A.

**Raman measurements at Notre Dame**

Room-temperature Raman spectroscopy measurements of the capped MnTe/GaAs thin film were performed using a Renishaw inVia InSpect Raman system with 532 nm laser excitation.

**Raman measurements at NIST**

Raman spectra were collected with a commercial HORIBA LabRAM HR Evolution confocal microscope using different excitation lasers. The spectra were collected in 180° backscattering geometry (1800 lines/mm grating; 100 µm confocal hole; 100× Olympus objective, 0.9 N.A.; wavelength-dependent spot size ≈ 1 µm). All measurements were performed at room temperature under ambient conditions unless otherwise stated. For power dependence measurements, the laser power is adjusted by adding different ND filters in the optical path and measured with a Thorlabs power meter after the objective and at the focus. To avoid degradation for some measurements, the samples were kept inside a commercial Linkam HFS600E microscopy cryostat under $1\times10^{-3}$ Torr, and in this case, the objective used was a 50× long working distance (Olympus, 0.5 N.A.).

To achieve angle-resolved Raman measurements, the HORIBA LabRAM HR Evolution system is customized, placing polarization optics that include a set of linear polarizers (03 FPG 003 Melles Griot Dichroic Sheet Polarizers) (P1, P2) and two super achromatic half-wave plates (Thorlabs, SAHWP05M-700) (HWP1, HWP2) for linear polarization. The waveplates are mounted in a custom-built motorized mount with a minimum step size of 0.5 °. Commercial Raman systems frequently utilize dichroic edge filters; consequently, the placement of the polarization optics is critical [52]. In this system, HWP1 is placed after the edge filter and before the microscope objective, and the HWP2 is placed after the edge filter, followed by the second polarizer, before the spectrograph.

**Raman measurements at George Mason**

Room-temperature Raman measurements were carried out in vacuum on bulk MnTe crystals in a backscattering geometry using a polarized 532 nm excitation laser focused onto the sample through a 40× objective (0.6 N.A.). These crystals had already been exposed to air for an extended duration. The incident laser power was maintained at approximately 550 µW, measured before the objective. Raman spectra were acquired with an integration time of 900 s per scan and averaged over 30 acquisitions to improve the signal-to-noise ratio. The scattered light was collected through the same objective and sent through three Bragg filters (Optigrate), which enabled measurements down to 15 $cm^{-1}$. The filtered light was then directed to a spectrometer equipped with a liquid-nitrogen-cooled charge-coupled device (CCD) detector.

**Cross-sectional STEM-EDS sample preparation at Northeastern**

To examine the depth-dependent oxidation morphology, a cross-sectional lamella was prepared, using focused ion beam (FIB) milling after the MnTe film had been exposed to ambient air for several weeks. Before trenching and thinning, a protective Pt cap was deposited over the air-exposed surface to preserve the pre-existing MnTe/$MnO_x$ interface and minimize $Ga^+$-induced sputtering, redeposition, curtaining, and structural modification during lamella preparation. The capped section was subsequently lifted out, attached to a TEM grid, and thinned to electron transparency before STEM-EDS analysis. Because the FIB and TEM measurements were performed in separate instruments, the prepared lamella underwent an unavoidable ambient-transfer and storage interval before being loaded into the TEM vacuum. Such ex-situ transfer can introduce surface contamination or additional oxidation of the newly exposed lamella faces. Nevertheless, the Pt cap protected the original top surface, and the strong localization of the O signal within the Mn-rich layer directly beneath the Pt, rather than uniformly throughout the MnTe film, indicates that the principal $MnO_x$ layer was already present before FIB preparation. Therefore, although limited oxidation of the exposed lamella faces during transfer cannot be excluded, it cannot account for the strongly localized Mn-O-rich layer observed at the original air-exposed surface.

**Supporting Information**

The Supporting Information includes, theoretical study of vibrational properties of MnTe, angle-resolved Raman spectroscopy of air-exposed MnTe/ GaAs films, structural characterization of epitaxial α-MnTe (0001)/ GaAs (111), elemental Te reference measurements, Raman signatures of surface $MnO_x$ formation, anti-Stokes Raman spectra, excitation-wavelength-dependent Raman measurements, and single-crystal magnetic susceptibility.

**Acknowledgments**

A.d.l.T. acknowledges helpful conversations with I. Mazin. S.K. and A.B. acknowledge support by the National Science Foundation through the Expand-QISE award NSF-OMA-2329067 and the Massachusetts Technology Collaborative award MTC-22032, and the computational resources of Northeastern University's Advanced Scientific Computation Center and the Explorer Cluster. S.B., X.L., and B.A.A. acknowledge support by the Air Force Office of Scientific Research under award number FA9550-25-1-0319. O.Y. gratefully acknowledges financial support for this publication by the Fulbright U.S. Scholar Program, which is sponsored by the U.S. Department of State. Its contents are solely the responsibility of the author and do not necessarily represent the official views of the Fulbright Program and the Government of the United States. O.Y. also acknowledges support from the National Research Foundation of Ukraine for financial support of this research (Project No. 2025.07/0271), "Band Structure Engineering in Mn- and Fe-Doped Multicomponent Dilute Ferromagnetic Semiconductors and Topological Semimetals for Photonic and Spintronic Device Applications." N.G. and R.B.R. were supported by the Army Research Office under Cooperative Agreement Number W911NF22-2-0173. T.H.Y.P., D.B., and P.M.V. acknowledge support from the National Science Foundation under grant no. 2226097. Commercial equipment, instruments, and materials are identified in this paper to adequately specify the experimental procedure. Such identification is not intended to imply recommendation or endorsement by the National Institute of Standards and Technology or the United States Government, nor is it intended to imply that the materials or equipment identified are necessarily the best available for the purpose.

This document has not been peer-reviewed but has been cleared by NIST for release.

**Conflict of Interest**

The authors have no conflicts to disclose.

**Data Availability Statement**

The data that support the findings of this study are available within the article and its supporting information.

## Supporting Information

# The Origin of the Observed Raman Peaks in α-MnTe

Nurul Azam,[1,2,†] Syed Mohammad Shahed,[1,2] Liam T. Schmidt,[1,2] Sara Bey,[3] Oksana Yastrubchak,[4] Maria F. Munoz,[5] Riccardo Torsi,[5] Thi Hai Yen Pham,[6,7] Dushyanthini Balasundaram,[6,7] Resham Babu Regami,[3,10] Wentao Liang,[8] Imrankhan Mulani,[1,2,9] Matthew Matzelle,[1,2] Vineet Kumar Sharma,[9] Sougata Mardanya,[9] Sugata Chowdhury,[9] Nirmal Ghimire,[3,10] Patrick M. Vora,[6,7] Angela R. Hight Walker,[5] Xinyu Liu,[3] Badih A. Assaf,[3] Arun Bansil,[1,2] Alberto De la Torre,[1,2] Swastik Kar[1,2,11,‡]

[1]*Department of Physics, Northeastern University, Boston, MA 02115, USA.*
[2]*Quantum Materials and Sensing Institute, Northeastern University Innovation Campus, Burlington, MA 01803, USA.*
[3]*Department of Physics and Astronomy, University of Notre Dame, Notre Dame IN, 46556, USA.*
[4]*V. E. Lashkaryov Institute of Semiconductor Physics, National Academy of Sciences of Ukraine, U-03028 Kyiv, Ukraine.*
[5] *Quantum Measurement Division, National Institute of Standards and Technology, Gaithersburg, MD, USA*
[6]*Department of Physics and Astronomy, George Mason University, Fairfax, VA 22030, USA*
[7]*Quantum Science and Engineering Center, George Mason University, Fairfax, VA 22030, USA*
[8]*Kostas Advanced Nano-Characterization Facility, Northeastern University Innovation Campus, Burlington, MA, USA.*
[9]*Department of Physics and Astrophysics, Howard University, Washington, DC 20059, USA.*
[10]*Stavropoulos Center for Quantum Matter, University of Notre Dame, Notre Dame IN, 46556, USA.*
[11]*Department of Chemical Engineering, Northeastern University, Boston 02115, USA.*

*Corresponding authors:* [†]*Nurul Azam (m.azam@northeastern.edu),* [‡]*Swastik Kar (s.kar@northeastern.edu)*

### 1.1 Vibrational properties of α-MnTe

The well-known altermagnetic α-MnTe has a hexagonal structure, crystallizing in space group $P6_3/mmc$ (194), featuring separate layers of Mn and Te atoms. Although numerous theoretical and experimental studies [1],[2],[3],[4] have sought to accurately characterize its vibrational properties, the existence of Raman-active modes remains a subject of debate. Therefore, using density functional perturbation theory [5],[6],[7] and symmetry analysis, we determine the Raman-active modes in this system and analyze how temperature-induced strain affects the Raman shift. To simulate the vibrational properties of α-MnTe, we used the experimental lattice constants of the bulk sample grown at 300 K, while the temperature effect was simulated by selecting lattice parameters for 100 K and 300 K MnTe grown on a GaAs substrate (see SI Figure S3), as shown in Table S1 [8]. Our calculations incorporate collinear antiferromagnetic ordering at the Mn sites, along with a Hubbard U correction with U = 4 eV and J = 0.97 eV.

The calculated phonon dispersion for MnTe along the high-symmetry path is shown in Figure S1(a). To determine the distinctive characteristics of these phonon modes, including Raman and infrared (IR) activity, a group-theory analysis is performed using the Phonopy package [7]. We determined that α-MnTe belongs to the point group $D_{6h}$ (6/mmm) and among the twelve phonon modes, the $E_{2g}$ mode at 94.81 $cm^{-1}$ is only Raman active, whereas the $A_{2u}$ mode at 130.42 $cm^{-1}$ and the $E_{1u}$ mode at 135.23 $cm^{-1}$ are IR active, as detailed in the table in Figure S1(b)**.**

Furthermore, growing α-MnTe on a GaAs substrate can induce in-plane tensile strain, which maintains a positive Poisson ratio. Our calculations show that this strain causes a downward shift of the predicted Raman mode in the vibrational spectrum. Specifically, the Raman frequency drops from 94.81 $cm^{-1}$ in the bulk sample to 93.74 $cm^{-1}$ in the GaAs-grown sample at 300 K. Additionally, raising the temperature of the MnTe-GaAs sample from 100 K to 300 K induces thermal expansion that results in tensile strain in both the in-plane and out-of-plane directions, further reducing the Raman frequency from 96.89 $cm^{-1}$ to 93.74 $cm^{-1}$, as shown in Figure S1(c)-(d).

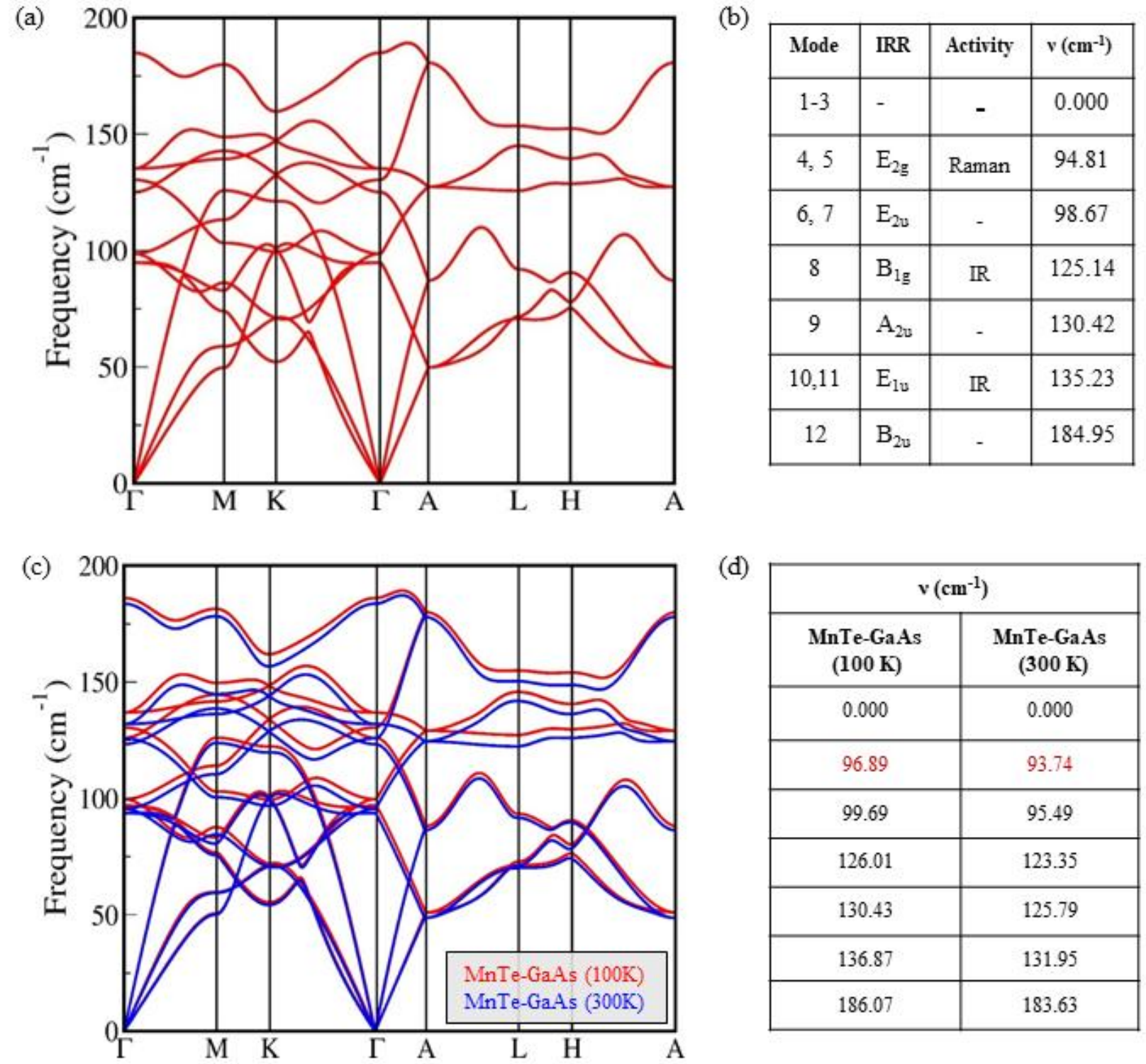

| Mode | IRR | Activity | v ($cm^{-1}$) |
|---|---|---|---|
| 1-3 | - | - | 0.000 |
| 4, 5 | $E_{2g}$ | Raman | 94.81 |
| 6, 7 | $E_{2u}$ | - | 98.67 |
| 8 | $B_{1g}$ | IR | 125.14 |
| 9 | $A_{2u}$ | - | 130.42 |
| 10,11 | $E_{1u}$ | IR | 135.23 |
| 12 | $B_{2u}$ | - | 184.95 |

| v ($cm^{-1}$) | |
|---|---|
| MnTe-GaAs (100 K) | MnTe-GaAs (300 K) |
| 0.000 | 0.000 |
| 96.89 | 93.74 |
| 99.69 | 95.49 |
| 126.01 | 123.35 |
| 130.43 | 125.79 |
| 136.87 | 131.95 |
| 186.07 | 183.63 |

**Figure S1.** Vibrational properties of MnTe include (a) phonon band dispersion along high-symmetry paths, (b) calculated phonon frequencies and mode activities, featuring one Raman-active and two IR-active modes. (c) Phonon dispersion for a MnTe film grown on GaAs substrate at 100 K and 300 K. Corresponding mode frequencies are tabulated in (d). Raman-active mode is highlighted in red color.

**Table S1.** Lattice parameters for different input structures of α-MnTe

| Materials | Lattice parameters | | |
|---|---|---|---|
| | a (Å) | c (Å) | c/a |
| Bulk (300K) | 4.1479 | 6.7118 | 1.618 |
| MnTe-GaAs (300 K) | 4.1780 | 6.6901 | 1.601 |
| MnTe-GaAs (100 K) | 4.1702 | 6.6204 | 1.588 |

## 1.2 Angle-Resolved Raman Spectroscopy of air exposed MnTe/GaAs films

Figure S2 shows the angular dependence of the Raman intensity of air exposed α-MnTe grown on GaAs. The polarization angle is controlled by the second half-wave plate (HWP2) placed after the edge filter as described in [9]. The observed angular dependence confirms the assignment of the near 90(5) $cm^{-1}$,120(3) $cm^{-1}$ and 140(3) $cm^{-1}$ peaks as originating from $E_1$, $A_1$ and $E_2$ Te modes [10].

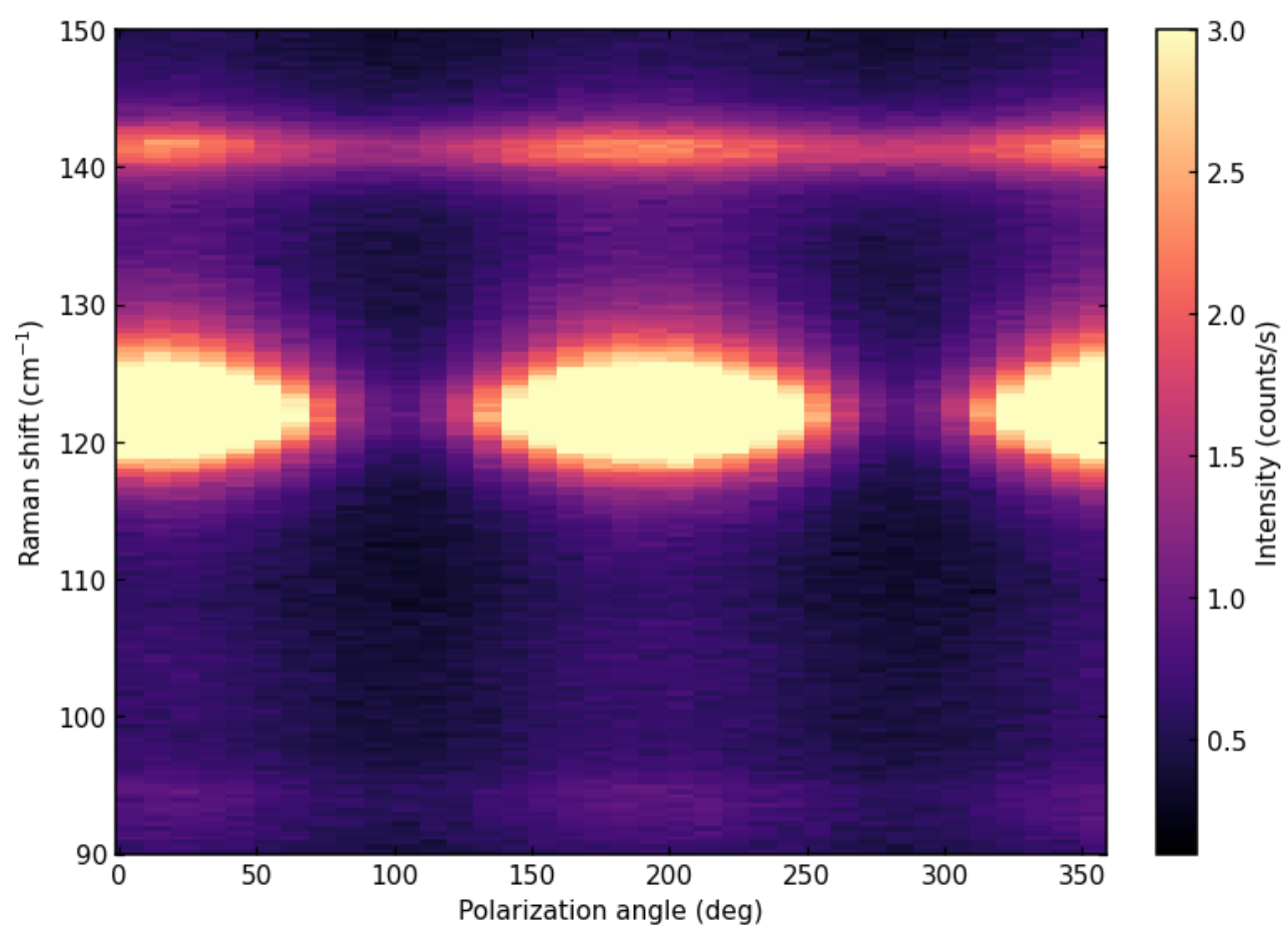

**Figure S2.** Angle-resolved Raman spectroscopy of air-exposed MnTe/GaAs.

## 1.3 Structural characterization of epitaxial α-MnTe (0001)/ GaAs (111)

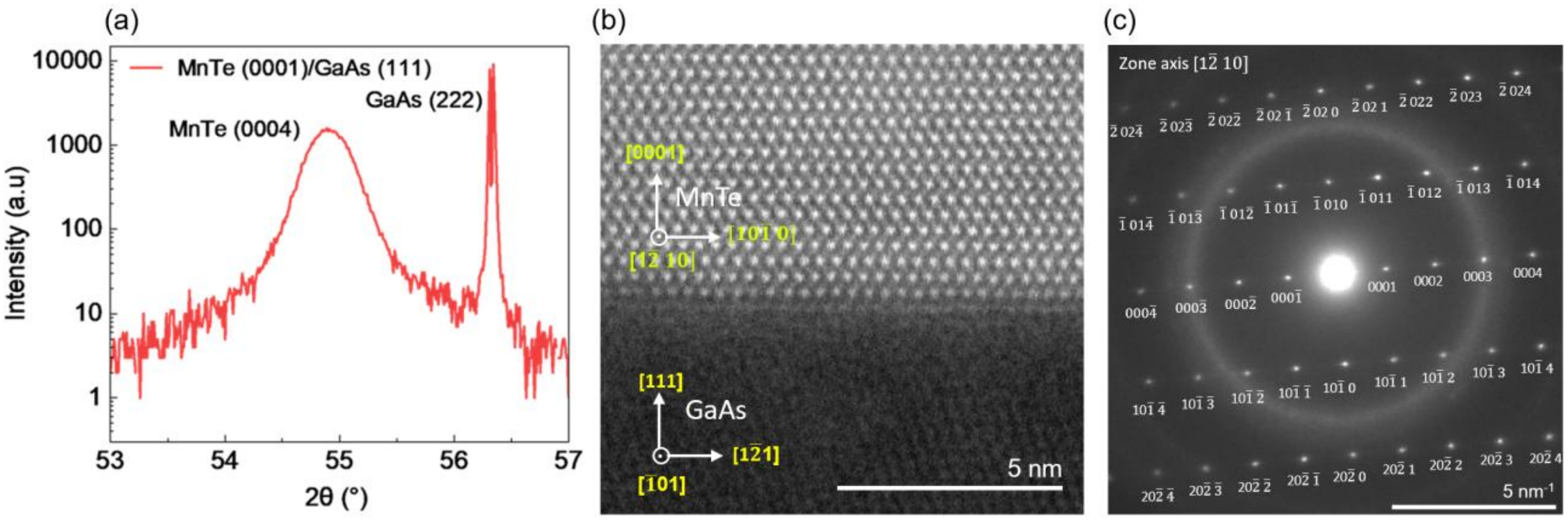

**Figure S3. (a)** X-ray diffraction scan showing the MnTe (0004) film reflection at $2\theta \approx 54.9°$ (bulk reference $2\theta = 54.60°$, PDF 01-090-6621) and the GaAs (222) substrate reflection at $2\theta \approx 56.35°$(PDF 00-032-0389), yielding an out-of-plane MnTe lattice parameter $c \approx 6.68$Å. **(b)** Cross-sectional HAADF-STEM image of the MnTe/GaAs interface, showing the epitaxial relationship $[0001]_{\mathrm{MnTe}} \parallel [111]_{\mathrm{GaAs}}$ (out-of-plane), $[10\bar{1}0]_{\mathrm{MnTe}} \parallel [1\bar{2}1]_{\mathrm{GaAs}}$ (in-plane), and $[1\bar{2}10]_{\mathrm{MnTe}} \parallel [\bar{1}01]_{\mathrm{GaAs}}$ (electron-beam direction). **(c)** Selected area electron diffraction (SAED) pattern acquired along the MnTe $[1\bar{2}10]$zone axis. The discrete diffraction spots are indexed consistently with hexagonal α-MnTe ($P6_3/mmc$), while the superimposed Debye-Scherrer rings originate from the polycrystalline Pt protective layer deposited during focused-ion-beam preparation of the cross-sectional TEM lamella and are not associated with the MnTe lattice. For clarity, the SAED pattern is cropped to display the $000l$ reflection series through $|\ l\ | = 4$; higher-order reflections lie outside the displayed field of view. The odd-$l$ (0001) and (0003) reflections are kinematically forbidden by the $000l$: $l = 2n$ systematic-absence condition imposed by the $6_3$screw axis, but acquire intensity through dynamical electron scattering, including double diffraction [11]. The measured reciprocal-lattice spacings yield $a \approx 4.18$ Å and $c \approx 6.70$ Å, consistent with the $c$-axis lattice parameter independently obtained from XRD.

## 1.4 XPS in elemental Te

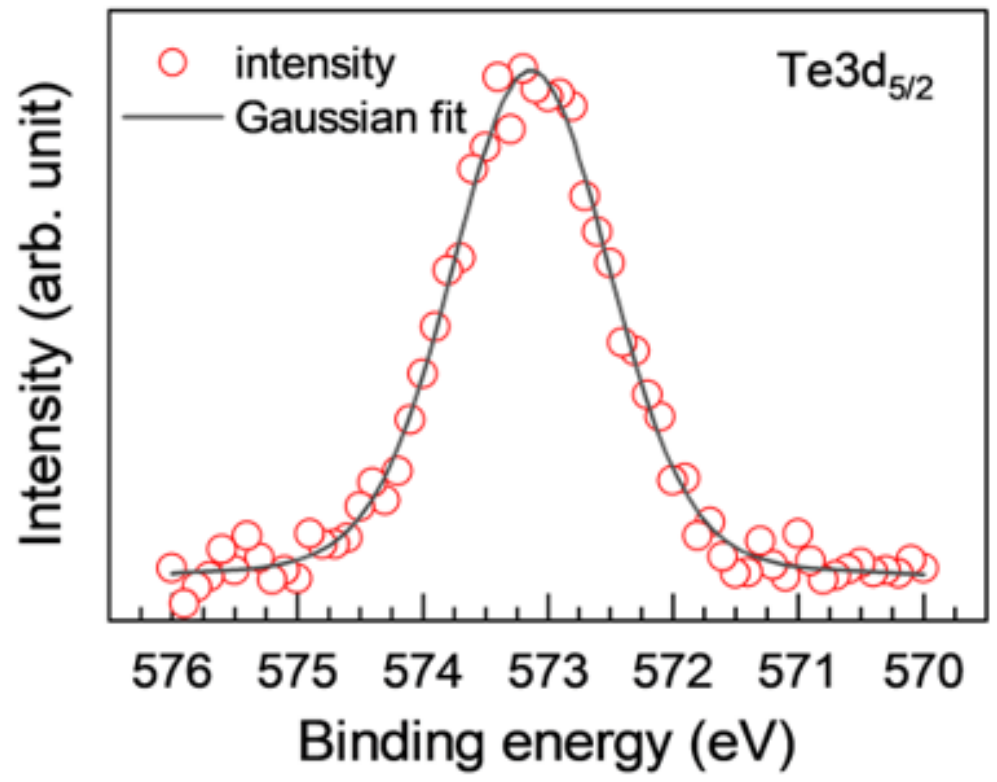

**Figure S4.** XPS reference spectrum of elemental Te deposited on GaAs. To establish the binding-energy position of elemental Te ($Te^0$), XPS was performed on a Te/GaAs reference sample. The Te $3d_{5/2}$ peak is centered near ≈573.0 eV, consistent with elemental Te. The open circles show the experimental data, and the solid line shows the Gaussian fit. This reference is used to support the assignment of the $Te^0$ component observed during air-induced oxidation of MnTe.

## 1.5 Raman signatures of surface $MnO_x$ formation in air-exposed MnTe

In addition to the Te-related low-frequency modes, two weak but reproducible Raman features appear near 506 $cm^{-1}$and 537 $cm^{-1}$ after air exposure. These features are absent in both the capped MnTe film and the elemental Te reference, indicating that they are associated with oxidation of MnTe rather than elemental Te or the intrinsic MnTe lattice. The observed peaks at ≈506 $cm^{-1}$ and ≈537 $cm^{-1}$ fall within the characteristic Mn-O vibrational range reported for manganese oxides, where modes near 500 $cm^{-1}$-515 $cm^{-1}$ and a broad MnO-related response around ≈530 $cm^{-1}$ have been reported [12], [13], [14]. The peaks near 267 $cm^{-1}$and 292 $cm^{-1}$ observed in all three spectra originate from the GaAs substrate. Together with the Mn-oxide layer independently identified by XPS and STEM-EDS, the 506 $cm^{-1}$ and 537 $cm^{-1}$ features provide additional spectroscopic evidence for surface $MnO_x$ formation during air exposure.

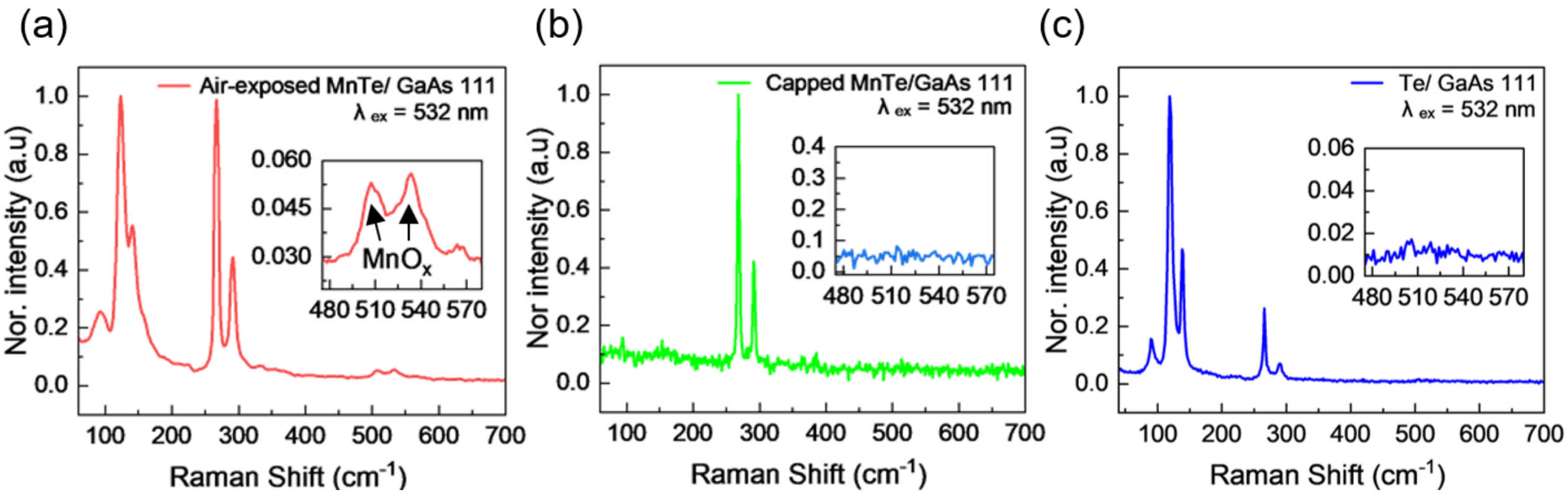

**Figure S5**. Raman signature of surface $MnO_x$ in air-exposed MnTe. Normalized Raman spectra ($\lambda_{ex}$ = 532 nm) of (a) air-exposed MnTe (0001)/ GaAs (111), (b) capped MnTe (0001)/ GaAs (111), and (c) an elemental Te/ GaAs (111) reference. Weak features near 506 $cm^{-1}$and 537 $cm^{-1}$ are observed only in the air-exposed MnTe, as highlighted in the inset, and are absent in both the capped MnTe and Te reference, supporting their assignment to surface $MnO_x$ formed during air-induced oxidation. [12], [13], [14] Peaks near 267 $cm^{-1}$and 292 $cm^{-1}$ present in all three spectra originate from the GaAs substrate.

## 1.6 Anti-Stokes Raman spectrum of α-MnTe/ GaAs film as a function of air exposure

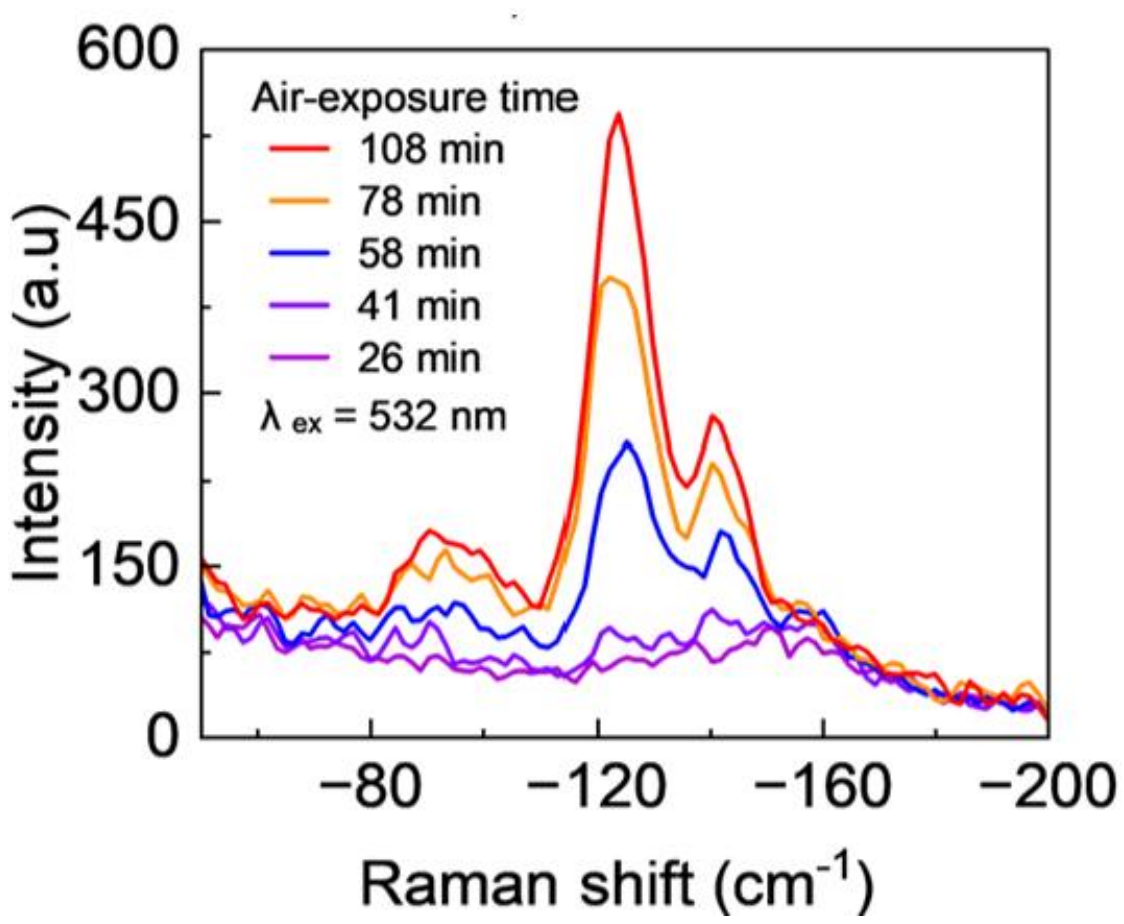

**Figure S6.** Time-dependent Anti-Stokes Raman spectra of MnTe during ambient-air exposure. This panel is the anti-Stokes counterpart of the Stokes spectra presented in Figure 5a. The modes near −120(3) $cm^{-1}$ and −140(3) $cm^{-1}$ progressively intensify between 26 min and 108 min, confirming the corresponding Stokes peaks and supporting their assignment as genuine Raman-active modes associated with Te segregation during MnTe degradation.

## 1.7 α-MnTe / GaAs Raman at different laser excitation wavelengths

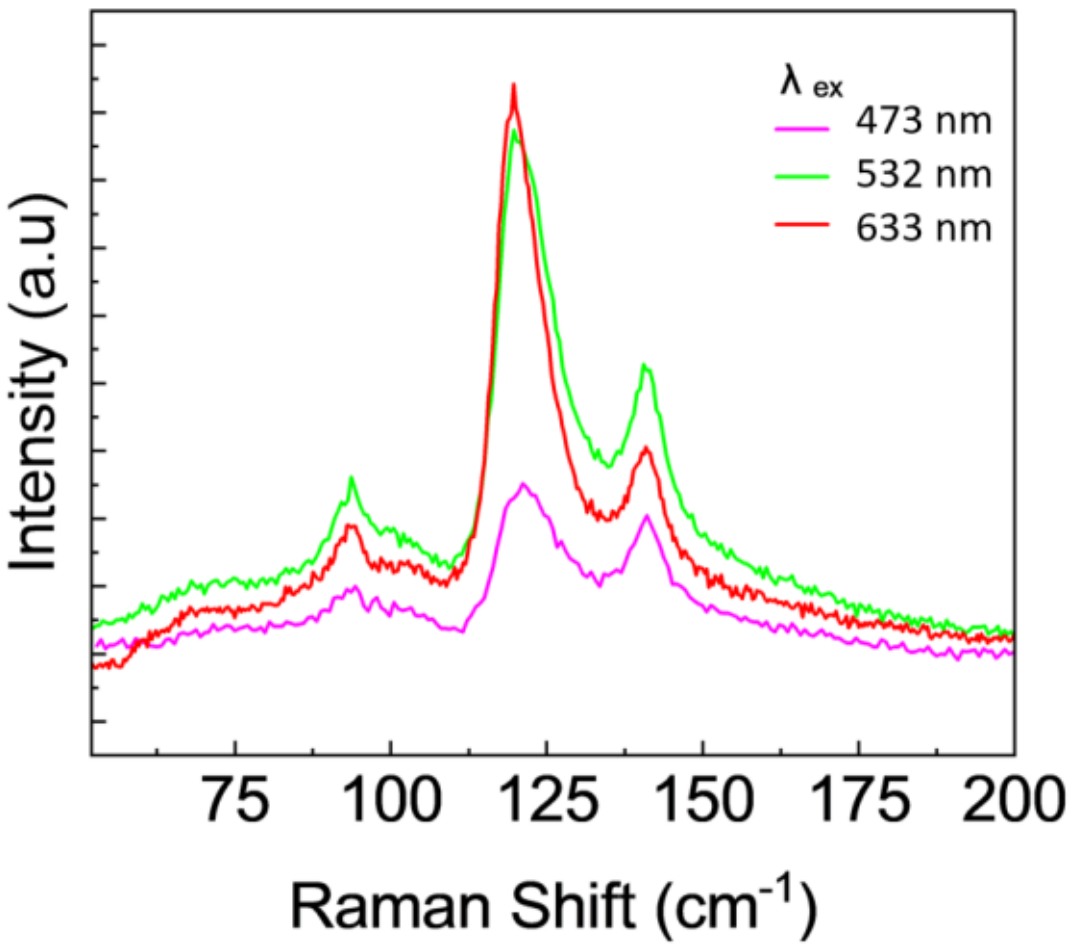

**Figure S7. Raman spectra of α-MnTe collected at 473 nm, 532 nm, and 633 nm excitation wavelengths**. The spectral features near 90(5) $cm^{-1}$, 120(3) $cm^{-1}$, and 140(3) $cm^{-1}$ are reproducibly observed under all three laser excitation conditions, while the relative intensities vary with excitation wavelength. The persistence of the ~120(3) $cm^{-1}$ and ~140(3) $cm^{-1}$ bands across different laser wavelengths confirms that these features are not specific to a particular excitation energy.

### 1.8 Magnetic susceptibility of the single crystal used in this study

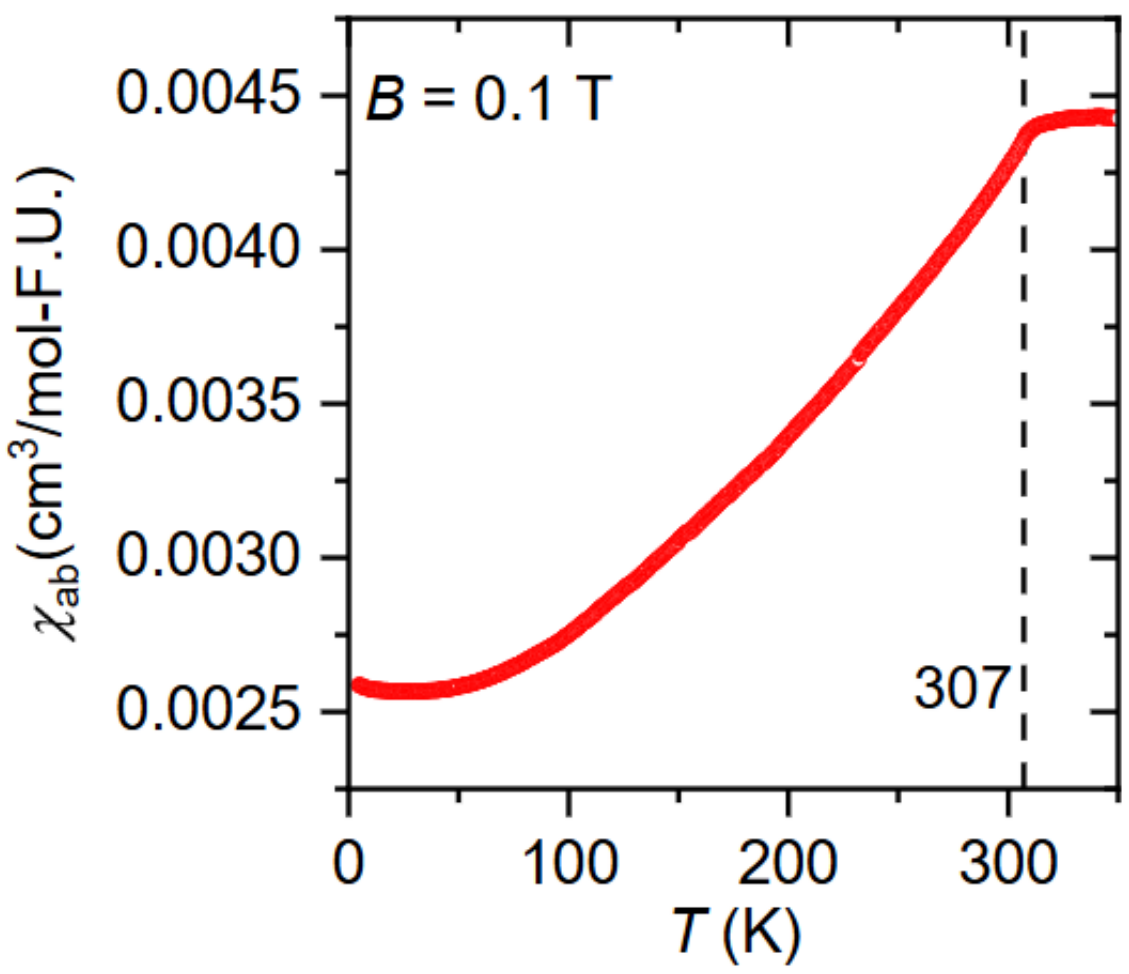

**Figure S8.** Temperature dependence of the magnetic susceptibility of MnTe single crystals used in this study with $T_N$ = 307 K, exemplifying the high sample quality.